\documentclass[reprint,aps,prd,superscriptaddress,preprintnumbers,nofootinbib]{revtex4-2}
\usepackage[utf8]{inputenc}
\usepackage[a4paper, total={7in, 9in}]{geometry}
\usepackage{amsmath}
\usepackage{amssymb}
\usepackage{enumerate}
\usepackage{amsmath}
\usepackage{tikz}
\usepackage[compat=1.1.0]{tikz-feynman}
\usepackage{feynmf}
\usepackage{slashed}
\usepackage{braket}
\usepackage{url}
\usepackage{natbib}
\usepackage{graphicx}
\usepackage[pdftex,bookmarks,linktocpage,pdfpagelabels,plainpages=false,hyperfigures,linkcolor=blue,citecolor=blue]{hyperref} 
\hypersetup{colorlinks=true}

\usepackage{mathrsfs,amssymb}  
\usepackage{cancel}
\usepackage[normalem]{ulem}

\usepackage{physics}
\usepackage{orcidlink}
\usepackage{appendix}

\begin{document}

\author{Rouzbeh Allahverdi \orcidlink{0000-0002-7134-3989}}
\email{rouzbeh@unm.edu}
\affiliation{Department of Physics and Astronomy, University of New Mexico, Albuquerque, NM 87131,
USA}

\author{Cash Hauptmann \orcidlink{0000-0003-3054-6189}}
\email{chauptmann2@huskers.unl.edu}
\affiliation{Department of Physics and Astronomy, University of Nebraska, Lincoln, NE 68588, USA}

\author{Peisi Huang \orcidlink{0000-0003-3360-2641}}
\email{peisi.huang@unl.edu}
\affiliation{Department of Physics and Astronomy, University of Nebraska, Lincoln, NE 68588, USA}

\title{Enhanced Dark Matter Abundance in First-Order Phase Transitions}

\begin{abstract}
    We propose a novel scenario to obtain the correct relic abundance for thermally under-produced dark matter. This scenario utilizes a strongly first-order phase transition at temperature $T_{\rm PT}$ that gives rise to dark matter mass $m$. Freeze-out in the broken phase can yield the desired abundance in the entire region currently allowed by observational bounds and theoretical constraints for $10^2 T_{\rm PT} \lesssim m \lesssim 10^4 T_{\rm PT}$. We show that the accompanying gravitational waves are strong enough to be detected by many upcoming and proposed experiments. This, in tandem with dark matter indirect searches, provides a multi-messenger probe of such models. Positive signals in the future can help reconstruct the potential governing the phase transition and shed light on an underlying particle physics realization.   
    
\end{abstract}

\maketitle

\section{Introduction}

There are various lines of evidence for the existence of dark matter (DM) in our universe~\cite{Bertone:2004pz}. However, the nature of DM remains a major open problem at the interface of cosmology and particle physics, and an important question concerns the relic abundance of DM that is precisely inferred from cosmological measurements~\cite{Komatsu:2014ioa, Planck:2013pxb,Planck:2018nkj,Planck:2018vyg}. An interesting and widely studied mechanism to address this question is thermal freeze-out in a radiation-dominated universe. The correct relic abundance can be obtained in this framework if the thermally averaged DM annihilation rate $\langle \sigma v \rangle$ takes the nominal value of 
$3 \times 10^{-26}$ cm$^3$ s$^{-1}$ at the time of freeze-out (coined ``WIMP miracle"). Indirect detection experiments have put $\langle \sigma v \rangle$ under increasing scrutiny. For example, Fermi-LAT’s
results from observations of dwarf spheroidal galaxies~\cite{Fermi-LAT:2015att}
and newly discovered Milky Way satellites~\cite{Fermi-LAT:2016uux} have placed limits that rule out thermal DM with a mass below 20~GeV in a model-independent way~\cite{Leane:2018kjk} (barring special cases with $p$-wave annihilation or coannihilation).

The correct relic abundance may be obtained for larger and smaller values of $\langle \sigma v \rangle$ if DM does not drop out of chemical equilibrium in a radiation-dominated universe. An important such scenario is DM production in nonstandard thermal histories with an epoch of early matter domination (EMD) prior to big bang nucleosynthesis (BBN)~\cite{Baer:2014eja}. In the case with $\langle \sigma v \rangle < 3 \times 10^{-26}$ cm$^3$ s$^{-1}$, which leads to thermal overproduction, entropy generation at the end of EMD can yield an acceptable relic abundance from freeze-out during EMD~\cite{Giudice:2000ex,Erickcek:2015jza,Nemevsek:2012cd,Nemevsek:2022anh,Nemevsek:2023yjl} or production at the end of EMD~\cite{Gelmini:2006pw,Allahverdi:2010rh}. In the thermally under-produced case, where $\langle \sigma v \rangle > 3 \times 10^{-26}$ cm$^3$ s$^{-1}$, the correct abundance can be obtained by a combination of DM production and its residual annihilation at the end of EMD~\cite{Kawasaki:1995cy,Moroi:1999zb}.     

Here we propose an alternative scenario that can yield the desired relic abundance for large values of $\langle \sigma v \rangle$. Instead of relying on a period of EMD, this scenario employs a first-order phase transition (FOPT) within the standard thermal history. The DM particles remain light at temperatures above that of the FOPT, denoted by $T_{\rm PT}$, and hence stay in chemical equilibrium with the thermal bath until then. The FOPT bumps the DM mass to values $m \gg T_{\rm PT}$, as a result of which DM particles undergo annihilation. For suitable values of $T_{\rm PT}$
one can obtain the correct DM relic abundance. Additionally, the FOPT is expected to form gravitational waves (GWs) that could be probed by upcoming detectors. This allows us to test this scenario via its GW signal in tandem with the current experimental bounds and future projections from indirect DM searches.

Our main goal is to examine viability of the proposed scenario and its observable signatures. Therefore we remain agnostic to the specifics of underlying particle physics realizations of this scenario. That said, our results for the GW signal fit well within models accommodating supercooled FOPTs such as classically conformal $B-L$ models~\cite{Iso:2009nw, Iso:2009ss, Jinno:2016knw, Marzo:2018nov,Huang:2022vkf}, Peccei-Quinn axion models~\cite{DelleRose:2019pgi, VonHarling:2019rgb}, strongly coupled theories~\cite{Baratella:2018pxi}, and extra dimensional theories~\cite{Agashe:2020lfz, Randall:2006py}. Moreover, an increase in DM mass after a FOPT may happen for any model with a fermionic DM candidate that is charged under a spontaneously broken chiral gauge symmetry~\cite{Abe:2023yte}.   





\section{Freeze-out and Abundance Enhancement}

The following Boltzmann equation provides the dynamics of the particle number (proper/physical) density $n$ of DM as it evolves in a Friedmann-Lemaître-Robertson-Walker spacetime:
\begin{equation}
    \label{eq:Boltzmann_equation}
    \dv{n}{t} = -3Hn - \langle \sigma v \rangle \left[ n^2 - n_\mathrm{eq}^2 \right] \, .
\end{equation}
Hubble's parameter $H \equiv (\dd a / \dd t) / a$ is given by the cosmic scale factor $a$ and its rate of change with respect to cosmic time $t$, $\sigma$ is the DM + DM annihilation cross section, and $v$ is the M{\o}ller speed between two interacting DM particles. Angular brackets denote a thermal average: $\langle \sigma v \rangle \equiv (\int \sigma v \, \dd n)/(\int \dd n)$. In deriving Eq.~\ref{eq:Boltzmann_equation} the DM is assumed to have a Maxwell-Boltzmann-like distribution function $f \propto e^{-E/T}$ in a Lorentz frame where $E^2 = m^2 + |\mathbf{p}|^2$ is the energy of DM with momentum $\mathbf{p}$, and $T$
is its temperature. The equilibrium density $n_\mathrm{eq}$ is given by an exact Maxwell-Boltzmann distribution:
\begin{equation}
    \label{eq:equilibrium_distribution}
    n_\mathrm{eq} = \int e^{-E/T} \frac{g}{(2\pi)^3} \dd ^3 p
\end{equation}
where $g$ is the number of internal degrees of freedom for the DM particle. This study uses $g=2$ for concreteness.

Given enough time, the cosmic expansion rate dominates the Boltzmann equation and DM densities are too sparse for number-changing processes to occur. This is called \textit{freeze-out} and happens at some temperature $T_f$. The final amount of DM can be characterized by its relic abundance $\Omega \equiv \rho_{\rm DM} / \rho_{\rm c,0}$; the ratio of DM energy density to today's critical energy density. Asymptotic solutions to the Boltzmann equation find
\begin{equation}
\label{eq:standard_relic_density}
    \Omega_{\rm thermal} = \frac{s_0}{\rho_{c,0}} \sqrt{\frac{45}{\pi}} \frac{\sqrt{g_*}}{g_{*s}} \frac{m}{M_\mathrm{Pl} T_f \langle \sigma v \rangle}
\end{equation}
(see Appendix \ref{appendix:freeze-out}). Today's entropy density is denoted with $s_0$, while $g_*$ and $g_{*s}$ respectively give the total relativistic degrees of freedom in the SM + DM plasma and entropy density~\cite{Borsanyi:2016ksw} in a radiation-dominated universe (we take $g_* = g_{*s} = 106.75 + 2$). The Planck mass is $M_{\rm Pl} = 1.22 \times 10^{19}$~GeV. This study assumes $s$-wave annihilation is dominant.

In the standard freeze-out analysis, Eq.~\ref{eq:standard_relic_density} is derived under the assumption that the DM number density closely tracks its equilibrium values for early times before freeze-out (called \textit{thermal} DM). Given sufficient interactions between DM and the SM, this assumption is justified for DM produced in these early times, as the particles will quickly reach thermal equilibrium with the dense universe.
However, if no appreciable number of massive DM particles exist 
for $T \geq T_f$ then Eq.~\ref{eq:standard_relic_density} cannot be applied. This has been studied in the context of non-thermal DM production from the decay of inflaton fields~\cite{PhysRevD.99.063508} or moduli fields in supersymmetric models~\cite{Fujii:2002kr, Acharya:2008bk, 10.1063/1.4883418}. For DM produced \textit{after} what would have been considered its characteristic freeze-out (had it been in equilibrium in the early universe), two cases must be considered, demarcated by some critical DM number density $n_c$
given by
\begin{equation}
    \label{eq:critical_density}
    n_c(T) = \frac{3H}{\langle \sigma v \rangle} \, ,
\end{equation}
which is defined for times after freeze-out ($T < T_f$) when the DM has fallen out of thermal equilibrium and therefore the equilibrium rate $\langle \sigma v \rangle n_\mathrm{eq}^2$ in Eq.~\ref{eq:Boltzmann_equation} is negligible. This critical number density is found by equating the remaining expansion and annihilation rates: $3Hn_c = \langle \sigma v \rangle n_c^2$.

If there is too little DM produced at $T_P < T_f$ and $n(T_P) < n_c(T_P)$, the cosmic expansion dominates annihilation. In this case, the relic abundance is already frozen in at $T_P$ and is determined by the DM abundance at its production: $\Omega_\mathrm{DM} = m \, n(T_P) / \rho_{c,0}$.
The more interesting case occurs when DM is produced in excess of its critical value in Eq.~\ref{eq:critical_density}.


When DM is produced at some temperature $T_P < T_f$ with density $n(T_P) > n_c(T_P)$, the annihilation rate dominates the cosmic expansion rate and rapid annihilation will occur until the two rates are equal and the DM density begins to stabilize. The same techniques of standard freeze-out analysis can be applied to find the subsequent Boltzmann evolution of this delayed, non-thermal production of DM. The resulting relic abundance finds an enhancement from the standard thermal DM scenario:
\begin{equation}
\label{eq:DM_relic_density_enhancement}
    \Omega_{\rm DM} = \Omega_{\rm thermal} \frac{T_f}{T_P}
\end{equation}
which is no longer a function of $T_f$ according to Eq.~\ref{eq:standard_relic_density}.
As long as the criterion $n(T_P) > n_c(T_P)$ is met, the exact value of the initial DM density at its production plays no role in determining the asymptotic values, as seen in Eq.~\ref{eq:DM_relic_density_enhancement}. Rapid annihilation drives the initial abundances toward a common attractor.

\section{Dark Matter Parameter Space}

\begin{figure*}
    \centering
    \includegraphics[scale=0.2]{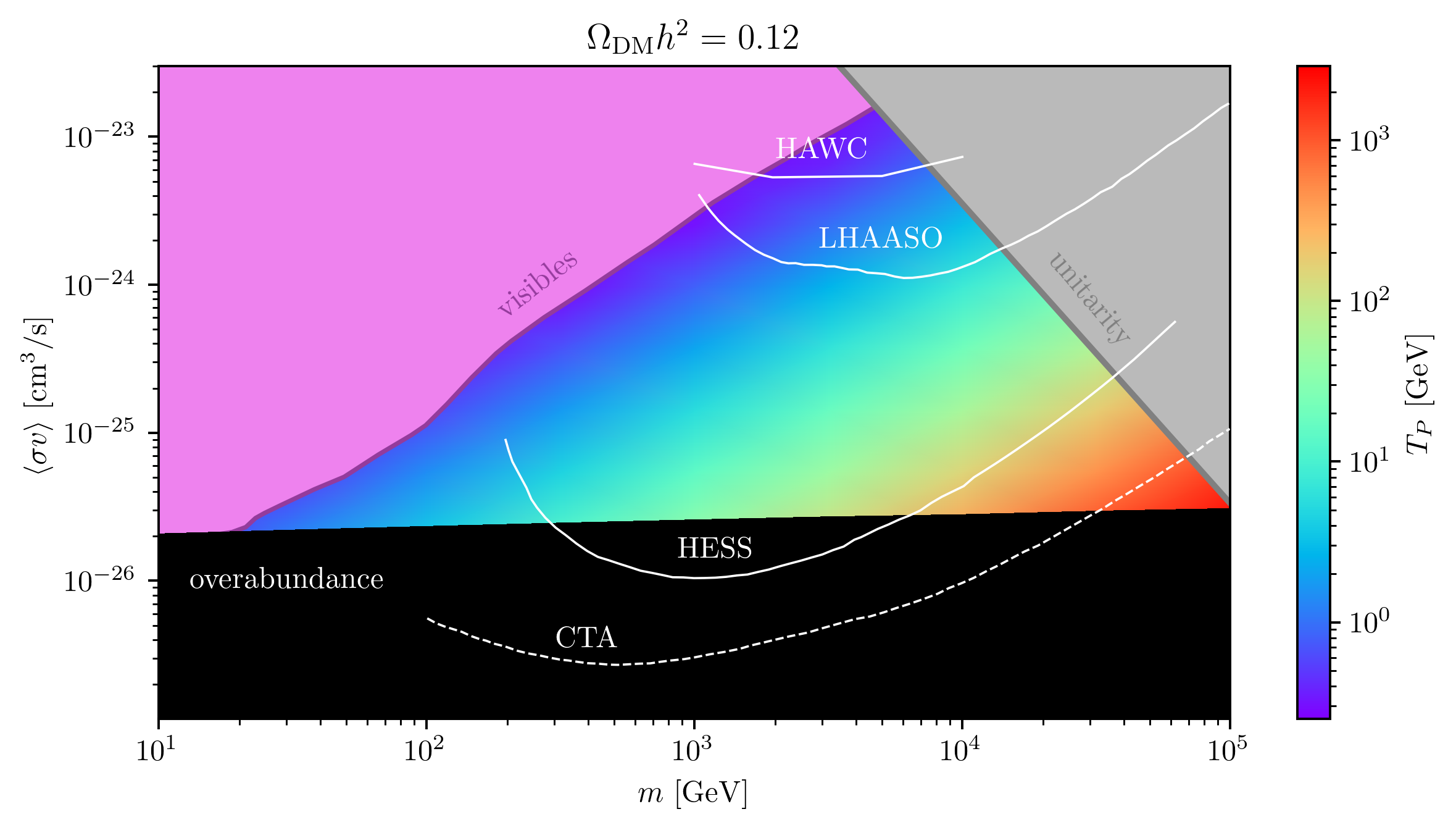}
    \caption{DM parameter space with cross section upper bounds on $s$-wave annihilation to visible final states~\cite{Leane:2018kjk} and unitarity~\cite{PhysRevLett.64.615}. Additional bounds are provided by searches of $\tau^- \tau^+$ DM annihilation channels from HAWC~\cite{HAWC:2023owv}, LHAASO~\cite{Bi:2022yyg}, and HESS~\cite{HESS:2022ygk}. The dashed curve provides the expected sensitivity of CTA~\cite{CTA:2020qlo} in the same channel. All $\tau^- \tau^+$ searches assume cusped DM density profiles in the Milky Way (Einasto profiles for HAWC, HESS, and CTA; Navarro-Frenk-White profile for LHAASO).}
    \label{fig:DM_parameter_space_production_temp}
\end{figure*}

For a given model of DM with mass $m$ and thermal cross section $\langle \sigma v \rangle$, Eq.~\ref{eq:DM_relic_density_enhancement} gives the necessary temperature $T_P$ at which the DM is produced to yield some relic abundance $\Omega_\mathrm{DM}$. The production temperatures which yield today's relic abundance of $\Omega_\mathrm{DM}h^2 = 0.12$~\cite{Planck:2018vyg} are shown in Fig.~\ref{fig:DM_parameter_space_production_temp}. Much of the DM parameter space has been ruled out by both theoretical and observational bounds. On the theoretical side, non-composite DM must have upper bounds on $\langle \sigma v \rangle$ to maintain unitarity in $2 \rightarrow 2$ interactions. Following arguments first made in Ref.~\cite{PhysRevLett.64.615}, we use a unitarity bound of $\langle \sigma v \rangle \lesssim 4\pi / (m^2 \langle v \rangle)$ where $\langle v \rangle \approx \sqrt{6m/T_f}$ (see Appendix~\ref{appendix:Moller_speed}). As for observational bounds, Ref.~\cite{Leane:2018kjk} combines limits on DM annihilation coming from Planck satellite measurements of the cosmic microwave background~\cite{Planck:2015fie}, Fermi-LAT measurements of dwarf spheroidal galaxies in the Milky Way~\cite{Fermi-LAT:2015att, Fermi-LAT:2016uux}, and the Alpha Magnetic Spectrometer measurements of cosmic rays~\cite{PhysRevLett.113.121101, PhysRevLett.113.121102}. These bounds only consider ``visible'' final states consisting of photons, leptons, or hadrons. Lastly, we carve out the parameter space which yields an overabundance of DM; models within this region find freeze-out times $T_f$ giving $\Omega_\mathrm{DM} > 0.12$. As the delayed production of DM above its critical abundance can only lead to an increase in its density, this region remains excluded for such $T_P < T_f$ models.

With the remaining DM parameter space of Fig.~\ref{fig:DM_parameter_space_production_temp}, we include single-channel bounds from searches of DM annihilating to $\tau^- \tau^+$ states from HAWC~\cite{HAWC:2023owv}, LHAASO~\cite{Bi:2022yyg}, and HESS~\cite{HESS:2022ygk} as this channel consistently provides the most stringent bounds across different experiments. Projected sensitivities of CTA~\cite{CTA:2020qlo} are shown by the dashed curve. Note that a significant portion of the DM parameter space allowed by HESS is expected 
to be probed by CTA searches, yet masses near 20 GeV or 100 TeV 
will still remain out of reach.

\section{Massive Dark Matter Production in Phase Transitions}

We now outline a simple mechanism which can produce massive DM in a cosmological FOPT occurring after its characteristic freeze-out time. Cosmological phase transitions are present in any particle physics model exhibiting symmetry breaking from an 
underlying (gauge) group. As this happens, the particles of the universe transition from a phase of interactions described by the original symmetry of the full (gauge) group to a phase of interactions given by some 
spontaneously broken subgroup. 
FOPTs are characterized by the broken phase nucleating in pockets that are stochastically distributed throughout the symmetric phase. As the FOPT proceeds, 
bubbles of broken phase expand 
until the universe is completely enveloped by the broken phase.

In this mechanism of DM ``production", the DM exists prior to the phase transition, however a change in its mass occurs. The usual scenario begins with massless 
fermionic species $\chi$ (eventually the DM candidate) 
with a Yukawa 
coupling to some scalar field $\phi$:
\begin{equation}
\label{eq:mass_gain_Yukawa_Lagranian}
    \mathcal{L} \supset - y \phi^* \overline{\chi}_R \chi_L + {\rm h.c.} \, ,
\end{equation}
where $\overline{\chi}_R$ and $\chi_L$ denote the right and left chiral Weyl components of $\chi$ respectively. {Our calculations use $g=2$ and therefore apply to Majorana DM ($\overline{\chi}_R = \chi_L$).\footnote{DM made of Dirac fermions with $g=4$ does not significantly alter our results as most of the thermal degrees of freedom are supplied by the Standard Model bath: $g_{\rm SM} = \mathcal{O}(100)$ for most production temperatures in Fig.~\ref{fig:DM_parameter_space_production_temp}.}
After the phase transition
the scalar field develops an asymmetric vacuum expectation value (VEV) $\langle \phi \rangle$ resulting in a mass term $m = y\langle \phi \rangle$ for the DM.

The massive DM abundance in the true vacuum is determined by the amount of massless 
$\chi$ in the false vacuum that have enough momentum to overcome the mass gap across the bubble wall. Assuming thermal equilibrium at the time of the FOPT, the number density of massless 
$\chi$ is given by $n_\mathrm{fv}(T_\mathrm{PT}) = (3 / 4\pi^2) \zeta(3) \gamma_w g T_\mathrm{PT}^3$. Therefore, if some fraction $n / n_\mathrm{fv}$ of 
$\chi$ particles penetrates into the true vacuum, the condition for abundance enhancement reads $(n / n_\mathrm{fv}) n_\mathrm{fv} \geq n_c(T_\mathrm{PT})$, or
\begin{equation}
    \label{eq:abundance_lower_bound_mass_gain}
    \frac{n}{n_\mathrm{fv}} \langle \sigma v \rangle T_\mathrm{PT} \geq
    \sqrt{\frac{2}{3}} \frac{8 \pi^{5/2}}{\zeta(3) M_\mathrm{Pl}} \frac{\sqrt{g_*}}{g} \, .
\end{equation}
The situation can be easily extended to other massive DM production mechanisms in FOPTs, such as scalar decay outlined in Appendix~\ref{appendix:scalar_decay}.

One comment is in order at this point. 
In general, mass of the particle that mediates interactions of DM may also change during a FOPT leading to an increase in $\langle \sigma v \rangle$ (for example, see~\cite{Borah:2023sal, Adhikary:2024btd}).
However, this will not affect the final DM abundance in our scenario. This is because the number density of $\chi$ particles can acquire and track the equilibrium value from scatterings in the thermal bath even if $\langle \sigma v \rangle$ is much (but not extremely) smaller than $3 \times 10^{-26}$ cm$^3$ s$^{-1}$ at $T \gg T_{\rm PT}$. All that matters is that $\langle \sigma v \rangle$ has reached its final value when the phase transition ends.

\section{Penetrating Bubble Walls} \label{sec:penetrating_bubble_walls_wall_speed_vw}

Eq.~\ref{eq:abundance_lower_bound_mass_gain} provides a lower bound on the abundance of DM particles in the true vacuum required for abundance enhancement. However, not all particles can penetrate into the true vacuum through the expanding bubble walls. When particles acquire a larger mass upon entering the true vacuum due to a nonzero VEV, only false vacuum particles with energies (relative to a bubble wall) above the mass difference can penetrate. More accurately, for a particle with three-momentum $\mathbf{p}$ and mass $m_\mathrm{fv}$ in the false vacuum, it could penetrate only if $|\mathbf{p}|^2 \geq m_\mathrm{tv}^2 - m_\mathrm{fv}^2$ where $m_\mathrm{tv}$ is its mass in the true vacuum. We will take $m_\mathrm{fv} = 0$ in the following. To calculate the fraction of particles that successfully transition to true vacuum, we assume the false vacuum populations are relativistic and in thermal equilibrium with the matter content of the universe, and we find the flux of particles with enough momentum to penetrate the walls.

In the frame of reference of a bubble wall with relative speed $v_w$, a relativistic particle of spin $s$ is described by the following one-particle phase space distribution function when in equilibrium:
\begin{equation}
    \label{eq:distribution_function}
    f_\mathrm{fv} = \frac{1}{e^{[\gamma_w |\mathbf{p}| (1 - v_w \cos{\theta}) - \mu]/T} - (-1)^{2s}} \, ,
\end{equation}
where $\gamma_w = (1-v_w^2)^{-1/2}$ is the Lorentz factor, $\mu$ is the chemical potential, and $\pi - \theta$ is the angle between the particle's three-momentum $\mathbf{p}$ and the wall's local velocity. The number density of fermions in the false vacuum is then
\begin{equation}
    \label{eq:number_density_false_vacuum}
    \begin{split}
        n_\mathrm{fv} &= \int f_\mathrm{fv} \frac{g}{(2\pi)^3} \dd^3p \\
        &= \frac{g}{2 \pi^2} \gamma_w \left[ \frac{3}{2} \zeta(3) + \frac{\pi^2}{6} \frac{\mu}{T} + \ln(2) \frac{\mu^2}{T^2} \right] T^3
    \end{split}
\end{equation}
where $g$ is the particle's number of internal degrees of freedom and $\zeta$ is the Riemann zeta function. The flux density of these particles which can penetrate a bubble wall is given by
\begin{widetext}
\begin{equation}
    \label{eq:flux_density_false_vacuum_penetration}
    \begin{split}
        J &= \int \frac{|\mathbf{p}| \cos\theta}{|\mathbf{p}|} f_\mathrm{fv} \Theta(|\mathbf{p}| \cos\theta - m) \frac{g}{(2\pi)^3} \dd^3p \\
        &= \frac{g}{(2\pi)^2} \left[ \frac{1}{2T^2 (1-v_w)^2 \gamma_w^3} \left\{ \left( m^2 (1-v_w)^2 \gamma_w^2 - \mu^2 \right) \left( -i\pi - 2\tanh^{-1}\left[ 1+2e^{\frac{m(1-v_w)\gamma_w-\mu}{T}} \right] \right. \right. \right. \\
        &\quad\quad \left. \left. \left. + \ln\left[ 1+e^{\frac{-m(1-v_w)\gamma_w+\mu}{T}} \right] \right) - 2mT(1-v_w)\gamma_w \mathrm{Li}_2{\left( -e^{\frac{-m(1-v_w)\gamma_w+\mu}{T}} \right)} - 2T^2 \mathrm{Li}_3{\left( -e^{\frac{-m(1-v_w)\gamma_w+\mu}{T}} \right)} \right\} \right] T^3
    \end{split}
\end{equation}
\end{widetext}
where $\Theta$ is the Heaviside step function and $\mathrm{Li}_q$ is the polylogarithm function of order $q$. As $J$ is the number of particles per area per time that can penetrate bubble walls, the number density of particles that pass into the true vacuum is $n = J / v_w$.
One can now find the fraction of particles which penetrate into the true vacuum and acquire mass:
\begin{equation}
    \label{eq:penetration_fraction}
    \frac{n}{n_\mathrm{fv}} = \frac{J}{v_w n_\mathrm{fv}} \, .
\end{equation}

Note that, as shown in Fig.~\ref{fig:penetration_fraction}, this fraction is a function only of wall speed $v_w$ and $x \equiv m / T$ whenever $\mu = 0$, which holds for Majorana fermions whose particle number freely changes\footnote{For Dirac fermions, there must be equal populations of particles and antiparticles to ensure $\mu = 0$ in chemical and thermal equilibrium. Chemical equilibrium requires the chemical potentials of the two populations to sum to zero. With equal number densities, Eq.~\ref{eq:number_density_false_vacuum} requires equality of the two chemical potentials. Satisfying both of these conditions demands $\mu = 0$.}.
We use this relationship to find the minimum wall speed for a given FOPT to produce today's relic abundance of DM in the following way. For DM produced after its characteristic freeze-out, requiring $\Omega_\mathrm{DM} h^2 = 0.12$ allows Eq.~\ref{eq:DM_relic_density_enhancement} to find the necessary time $x_P = m / T_P$ at which the DM must be produced. Yet, the abundance enhancement only applies if Eq.~\ref{eq:abundance_lower_bound_mass_gain} is satisfied, requiring a minimum fraction of penetrating particles. Thus, a given $x_P$ and minimum $n / n_\mathrm{fv}$ can identify a minimum wall speed.
Choosing the \textit{minimum} penetrating fraction $n/n_{\rm fv}$ as outlined above is not necessary to the DM production mechanism---this fraction can take any value satisfying Eq.~\ref{eq:abundance_lower_bound_mass_gain} to overcome the critical DM number density. This is in contrast to mechanisms using ``filtered DM"~\cite{Chway:2019kft, Baker:2019ndr}, which rely in part on inefficient DM penetration.

We end this section with a discussion on the possibility of forming compact objects from DM trapped in the false vacuum. As a FOPT nears completion, bubbles of the true vacuum dominate the universe and domains or remnants of the false vacuum shrink as bubbles continue to expand. These remnants may contain those DM particles which lack the sufficient three-momenta to overcome the mass gap across bubble walls. If regions within these remnants become sufficiently dense, compact DM objects may form under their own gravitational interactions. For example, these compact objects could be black holes~\cite{Gross:2021qgx, Baker:2021nyl, Baker:2021sno} or, if the DM is fermionic, an intermediate stage of Fermi-ball formation~\cite{Kawana:2021tde, Hong:2020est,Huang:2022him,Xie:2024mxr} may take place before collapsing into black holes. Whatever the mechanism, DM annihilation can severely preclude the formation of compact objects. To keep DM from completely annihilating within false vacuum remnants, particle models must have either an asymmetric distribution of DM to anti-DM populations, or some suppression of the annihilation mechanism via small Yukawa couplings or kinematic constraints. Since our analysis has no requirement for small Yukawa couplings between the DM and the FOPT-inducing scalar field, we do not consider scenarios with suppressed DM annihilation, and therefore assume no appreciable formation of compact objects before DM annihilates within (or escapes from) domains of the false vacuum.


\begin{figure}
    \centering
    \includegraphics[scale=0.172]{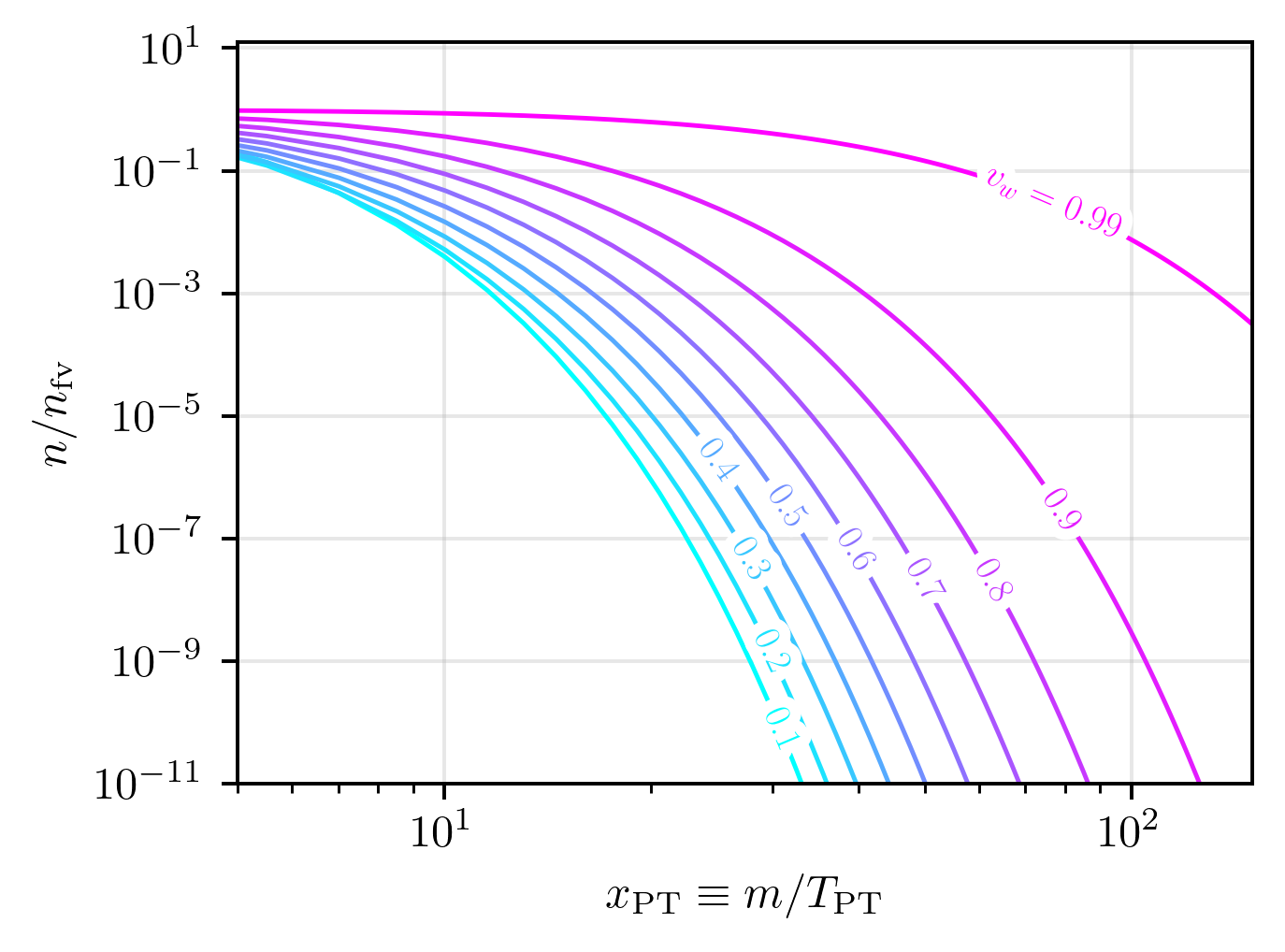}
    \caption{The fraction of particles able to penetrate into true vacuum and acquire mass for different bubble wall speeds (line labels). A chemical potential of $\mu = 0$ is assumed.}
    \label{fig:penetration_fraction}
\end{figure}

\section{Gravitational Waves}
\label{sec:gravitational_waves}

Particularly strong FOPTs could displace enough energy to produce a stochastic GW background detectable by future experiments which aim to measure the amplitudes and frequencies of GWs. FOPTs that release more latent heat produce GWs of greater amplitudes and are thus easier to detect. Given the stochastic nature of FOPTs, their GW signal is expected to yield a wide range of frequencies peaked around some frequency heavily influenced by the characteristic timescale of the phase transition. These two properties of a FOPT, its latent heat and timescale, are characterized by $\alpha$ and $\beta$:
\begin{equation}
    \label{eq:alpha_and_beta}
    \begin{split}
        \alpha &= \frac{1}{\rho_r(T)} \left\{ [V_\mathrm{eff}(\phi_\mathrm{false}, T) - V_\mathrm{eff}(\phi_\mathrm{true}, T)] \right. \\
        & \left. - T \left[ \dv{V_\mathrm{eff}(\phi_\mathrm{false}, T)}{T} - \dv{V_\mathrm{eff}(\phi_\mathrm{true}, T)}{T} \right] \right\} \bigg\rvert_{T=T_\mathrm{PT}} \\
        \beta &= - \dv{(S_3 / T)}{t} \bigg\rvert_{t=t_\mathrm{PT}}
        \, .
    \end{split}
\end{equation}
Here, $\rho_r(T) = \pi^2 g_*(T) T^4 / 30$ is the energy density of the radiation bath at the time of the FOPT with $g_*$ the relativistic degrees of freedom in the plasma. Subscripts ``true'' and ``false'' denote quantities evaluated in the true and false vacuua. More useful in GW calculations is $\beta / H(T_\mathrm{PT})$ where $H$ is the Hubble parameter. $T_\mathrm{PT}$ is a characteristic temperature at which the FOPT occurs. In what follows, we also take this to be the temperature at which massive DM is produced ($T_\mathrm{PT} = T_P$). It should be noted that $T_\mathrm{PT}$ is a post-supercooling temperature and all reheating effects take place before this time.

In a FOPT, the three dominant contributions to a GW background come from (1) sound waves in the universal plasma, (2) collisions of true vacuum bubbles, and (3) magnetohydrodynamic turbulence in the plasma. Following Ref.~\cite{Caprini:2015zlo}, we approximate the total GW background to be a linear combination of these stochastic sources:
\begin{equation}
    \label{eq:total_GW_spectrum}
    \Omega_\mathrm{GW} = \Omega_\mathrm{sw} + \Omega_\mathrm{col} + \Omega_\mathrm{turb}
\end{equation}
with $\Omega = \rho / \rho_c$ the corresponding fraction of the critical energy density, however GW signal spectra are usually presented in terms of $\Omega(f) \equiv \mathrm{d}\Omega / \mathrm{d}\ln(f)$~\cite{Maggiore:1999vm, PhysRevD.59.102001}, which is the notation used here. To determine each of these contributions, one needs knowledge of the fractional amount of available latent heat that transfers into (1) the bulk motion of the fluid ($\kappa_\mathrm{sw}$), (2) the bubble wall kinetic energy ($\kappa_\mathrm{col}$), and (3) turbulence ($\kappa_{\rm turb}$).

The sound wave contribution is numerically fitted as
\begin{equation}
    \label{eq:sound_wave_GW_contribution}
    \begin{split}
        h^2 \Omega_\mathrm{sw}(f) &= 3 \times 2.65 \times 10^{-6} \left[ \frac{H(T_\mathrm{PT})}{\beta} \right] \left[ \frac{\kappa_\mathrm{sw} \alpha}{1+\alpha} \right]^2 \\
        & \times \left[ \frac{100}{g_*(T_\mathrm{PT})} \right]^{1/3} v_w \left[ \frac{f}{f_\mathrm{sw}} \right]^3 \left[ \frac{7}{4 + 3(f/f_\mathrm{sw})^2} \right]^{7/2} \\
        & \times H(T_{\rm PT}) \tau_{\rm sw}, \\
        f_\mathrm{sw} &= \frac{1.15}{v_w} \left[ \frac{\beta}{H(T_\mathrm{PT})} \right] h_* \,\, , \\
        h_* &= 1.65 \times 10^{-5} \, \mathrm{Hz} \left[ \frac{T_\mathrm{PT}}{100 \, \mathrm{GeV}} \right] \left[ \frac{g_*(T_\mathrm{PT})}{100} \right]^{1/6}
    \end{split}
\end{equation}
where $\kappa_\mathrm{sw}$ is the fraction of latent heat transferred into the bulk motion of the plasma, and $g_*(T_\mathrm{PT})$ is the number of relativistic degrees of freedom for all fields when the FOPT occurs. (The prefactor of 3 is found in the erratum of Ref.~\cite{Hindmarsh:2017gnf}.) The effective number of degrees of freedom coming from relativistic SM particles changes with temperature~\cite{Borsanyi:2016ksw}. Redshifting the peak frequencies to their expected values today is accomplished with the $h_*$ factor which assumes adiabatic expansion of a radiation-dominated universe. For conservative estimates, a suppression factor of $H \tau_{\rm sw}$ is used. This factor accounts for the finite lifetime of sound waves as sources of GWs, and is given by~\cite{Ellis:2020awk, Guo:2020grp}
\begin{equation}
    \label{eq:shock_time_suppression}
    \begin{split}
    H \tau_{\rm sw} &= \min[1, H \tau_{\rm sh}] \, , \\
    H \tau_{\rm sh} &\approx
    (8\pi)^{1/3} v_w \left( \frac{\beta}{H} \right)^{-1} \left( \frac{3}{4} \frac{\kappa_{\rm sw} \alpha}{1+\alpha} \right)^{-1/2} \, .
    \end{split}
\end{equation}

A numerical fit for the contributions from magnetohydrodynamic turbulence is found to be~\cite{Ellis:2020awk}
\begin{equation}
    \label{eq:MHD_turbulence_GW_contribution}
    \begin{split}
        h^2 \Omega_\mathrm{turb}(f) &= 3.35 \times 10^{-4} \left[ \frac{H(T_\mathrm{PT})}{\beta} \right] \left[ \frac{\kappa_\mathrm{sw} \alpha}{1+\alpha} \right]^{3/2} \\
        & \times \left[ \frac{100}{g_*(T_\mathrm{PT})} \right]^{1/3} v_w \left[ 1 - H(T_{\rm PT}) \tau_{\rm sw} \right] \\
        & \times \left[ \frac{(f/f_\mathrm{turb})^3}{[1+(f/f_\mathrm{turb})]^{11/3} [1+8\pi f/h_*]} \right] , \\
        f_\mathrm{turb} &= \frac{1.64}{v_w} \left[ \frac{\beta}{H(T_\mathrm{PT})} \right] h_* \, .\\
    \end{split}
\end{equation}

Finally, for a strongly supercooled phase transition, we use lattice simulation results from Ref.~\cite{Lewicki:2020jiv} to provide a numerical fit for the contributions from collisions of bubble walls: 
\begin{equation}
    \label{eq:bubble_wall_collisions_GW_contribution}
    \begin{split}
        h^2 \Omega_\mathrm{col}(f) &= 1.67 \times 10^{-5} \left[ \frac{H(T_\mathrm{PT})}{\beta} \right]^2 \left[ \frac{\kappa_\mathrm{col} \alpha}{1+\alpha} \right]^2 \\
        & \times \left[ \frac{100}{g_*(T_\mathrm{PT})} \right]^{1/3} \left[ \frac{1 + \left(\frac{2\pi f}{0.13 \beta}\right)^{d-a}}{1 + \left(\frac{2\pi f_\mathrm{col}}{0.13 \beta}\right)^{d-a}} \right] \\
        & \times \frac{3.63 \times 10^{-2} \times (a+b)^{c}}{\left[b\left(f/f_\mathrm{col}\right)^{-a/c} + a\left(f/f_\mathrm{col}\right)^{b/c} \right]^{c}}, \\
        f_\mathrm{col} &= \frac{0.81}{2\pi} \left[ \frac{\beta}{H(T_\mathrm{PT})} \right] h_* \, .
    \end{split}
\end{equation}
Here, $\kappa_\mathrm{col}$ giving the fraction of latent heat converted into kinetic energy of the bubble walls. The numerical parameters are $a=2.54$, $b=2.24$, $c=2.30$, and $d=0.93$.

To meet relic abundance requirements, Fig.~\ref{fig:DM_parameter_space_production_temp} shows that the DM must be produced at times $x_P \equiv m/T_P \sim [10^2, 10^4]$. If DM mass is proportional to the scalar field's VEV, high $x_P$ indicates the FOPTs should occur at temperatures well below the VEV, given that 
$y \leq \mathcal{O}(1)$ in the perturbative regime. 
This suggests the FOPT is supercooled, meaning it completes at some temperature well below its critical temperature (the temperature, typically on the same energy scale of the VEV, at which the effective potential acquires degenerate minima). With supercooling present, at the end of the FOPT a large difference in the effective potential obtains between the true and false vacuum phases, releasing much more energy than a non-supercooled transition. For this analysis, we thus assume $\alpha \gg 1$ meaning the contributions to $\Omega_\mathrm{GW}(f)$ are practically independent of $\alpha$\footnote{For $\alpha \approx 1$, the GW signal strength is only reduced by a factor of about four. Even with this reduction, Fig.~\ref{fig:GW_peaks_mass_gain} shows GW signals are still well within projected detector sensitivities. On the other hand, $\alpha$ should not be allowed arbitrarily large values to avoid an extended period of vacuum energy domination, requiring adjustments to the GW signal calculations.}. In taking $\alpha \gg 1$ limits, we also calculate $\kappa_\mathrm{sw} = \kappa_\mathrm{sw}(\alpha, v_w)$ using numerical fits found in Ref.~\cite{Espinosa:2010hh}. For values of $v_w$ required in these fits and in $\Omega_\mathrm{GW}(f)$, we use the minimum wall speed required for abundance enhancement outlined in Sec.~\ref{sec:penetrating_bubble_walls_wall_speed_vw}. In the remaining GW spectra parameters $\kappa_\mathrm{col}$ and $\kappa_\mathrm{turb}$, simulations~\cite{Hindmarsh:2015qta} have shown power in transverse modes of plasma fluid velocity to be at most $5-10\%$ of that in longitudinal modes. We thus conservatively take $\kappa_\mathrm{turb} = 0.05 \kappa_\mathrm{sw}$. Finally, to find $\kappa_\mathrm{col}$ we enforce $\kappa_\mathrm{sw} + \kappa_\mathrm{turb} + \kappa_\mathrm{col} = 1$ which assumes all latent heat is dispersed among sound waves, turbulence, and bubble collisions.

\begin{figure*}
    \centering
    \includegraphics[scale=0.2]{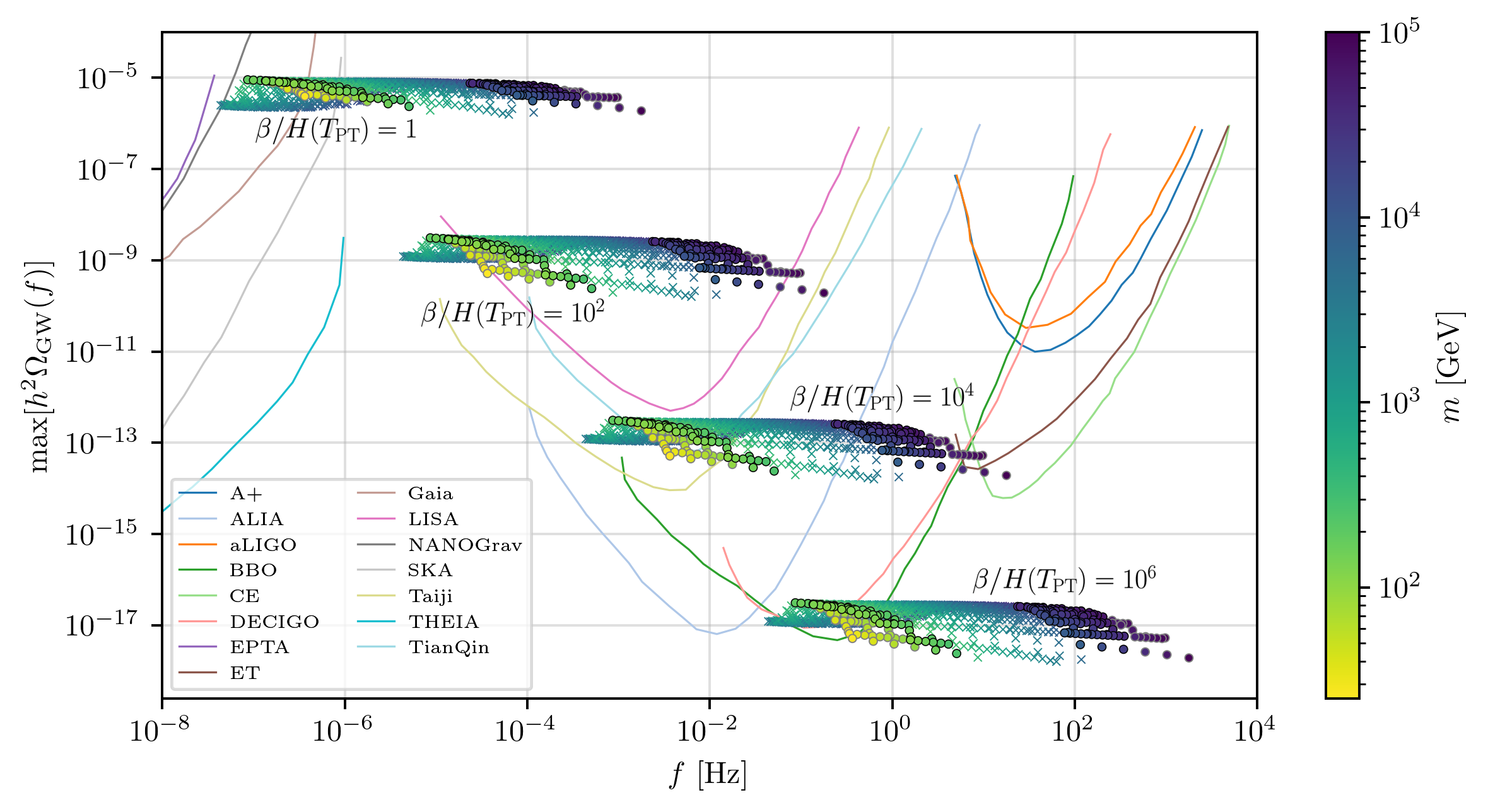}
    \caption{Peak values of GW signal for different $\beta / H(T_\mathrm{PT})$. Each peak is colored to its corresponding DM mass. Disks are points compatible with HESS bounds while crosses are not. Additionally, disks outlined in black are within CTA's projected sensitivities. Colored lines show the projected sensitivities of the following proposed experiments:
    A+~\cite{A+:LIGO}, ALIA~\cite{Gong:2014mca}, aLIGO~\cite{Shoemaker:2019bqt, LIGOScientific:2014pky}, BBO~\cite{Corbin:2005ny, Yagi:2011wg}, CE~\cite{LIGOScientific:2016wof}, DECIGO~\cite{Kawamura:2020pcg, Kawamura:2006up}, EPTA~\cite{EPTA:2011kjn}, ET~\cite{Punturo:2010zz}, Gaia~\cite{brown2018gaia}, LISA~\cite{Caprini:2019egz, amaro2017laser, Robson:2018ifk}, NANOGrav~\cite{NANOGrav:2020bcs, Lazio:2017fos}, SKA~\cite{Janssen:2014dka}, Taiji~\cite{Ruan:2018tsw}, THEIA~\cite{2018FrASS...5...11V}, TianQin~\cite{TianQin:2015yph}.}
    \label{fig:GW_peaks_mass_gain}
\end{figure*}

\section{Results}

To calculate GW spectra, we take points in the $m$-$\langle \sigma v \rangle$ plane, shown in Fig.~\ref{fig:DM_parameter_space_production_temp}, which yield the correct DM relic abundance. Furthermore, we do not consider the points excluded by visibles and unitarity bounds. Within this study, the only free parameter left in calculating GW spectra is $\beta / H(T_\mathrm{PT})$, which we take to be greater than one to ensure bubble nucleation rates exceed universal expansion rates and the FOPT can complete. Peak values of GW signals from FOPTs are plotted in Fig.~\ref{fig:GW_peaks_mass_gain} to compare against expected sensitivities of proposed GW detection experiments. It is clear that supercooled FOPTs with $\alpha \gg 1$ help provide GW signals strong enough to reach expected sensitivities for a wide range of $\beta / H(T_\mathrm{PT})$. Although the correlation is not exact, we can also see that in models endowing larger masses to DM fields, the resulting GW signals peak at higher frequencies; suggesting these models require more rapid FOPTs. As mentioned above, each signal peak is determined by a point in Fig.~\ref{fig:DM_parameter_space_production_temp} which is not excluded by bounds on visible annihilation channels, unitarity, and overabundance. The peaks are further distinguished by their shape: disks ($\circ$) are points permitted by HESS bounds while crosses ($\times$) are not. In addition to evading HESS bounds, disks with black outlines are probeable by projected sensitivities of CTA.

Finally, in Fig.~\ref{fig:multi-messenger_parameter_space} we offer a fuller look at possible DM candidates in a parameter space mixing particle characteristics and GW data from the accompanying FOPT. In this plot, the DM particle mass is given by the vertical axis while thermally averaged cross sections color each point. The horizontal axis provides the frequencies of peak GW signals, and we include four frequency bands demarcating the projected sensitivity ranges of THEIA, LISA, ALIA, and ET. Note that although a given point may fall within one of these frequency bands, its GW signal strength will determine whether the point might be probed by some detector, hence our omission of points with $\beta / H(T_{\rm PT}) = 10^6$ from Fig.~\ref{fig:multi-messenger_parameter_space}. As in Fig.~\ref{fig:GW_peaks_mass_gain}, disks denote points compatible with HESS bounds while crosses do not; disks with black outlines 
can be further probed by CTA. The three distinct (although overlapping) sets of points correspond to different $\beta / H(T_\mathrm{PT})$ values in the FOPTs, labeled near the bottom of the plot.

\begin{figure*}
    \centering
    \includegraphics[scale=0.2]{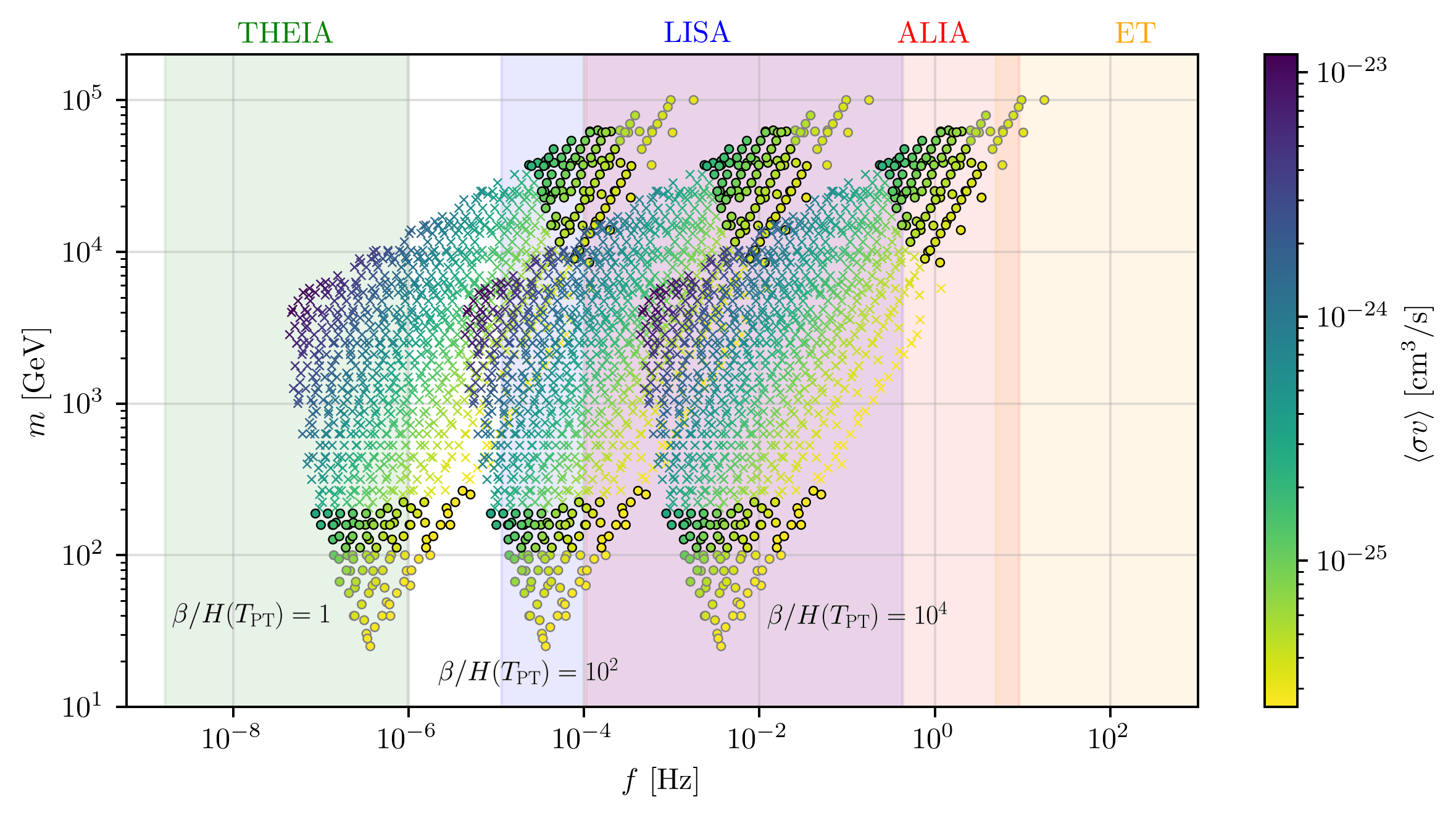}
    \caption{Mixed space of DM particle properties and GW parameters. Point shapes give the same designations as Fig.~\ref{fig:GW_peaks_mass_gain}.}
    \label{fig:multi-messenger_parameter_space}
\end{figure*}

\section{Conclusion}

In this paper, we presented a new scenario for obtaining the correct relic abundance in the case of thermally under-produced DM, i.e., $\langle \sigma v \rangle > 3 \times 10^{-26}$ cm$^3$ s$^{-1}$. This scenaio utilizes a FOPT in a radiation-dominated universe as a result of which DM acquires its mass $m$. DM particles maintain a constant comoving density above FOPT temperature $T_{\rm PT}$ and their freeze-out in the broken phase yields the correct abundance if $T_{\rm PT}$ is sufficiently lower than the freeze-out temperature in the standard scenario $T_{\rm f} \sim m/20$. 

In particular, we found that $10^2 T_{\rm PT} \lesssim m \lesssim 10^4 T_{\rm PT}$ within the mass range allowed by current experimental limits from indirect searches and theoretical bounds from unitarity. The hierarchy between $m$ and $T_{\rm P}$ suggests that the FOPT must be supercooled thereby leading to potentially significant GW signals. We found that although the GWs sourced by such a FOPT would likely not be detectable by current experiments, they are within projected sensitivities of many next-generation GW detectors. 

As shown in Fig.~\ref{fig:GW_peaks_mass_gain}, for reasonable values of PT parameters, the GW signal is within the reach of experiments like ET, LISA, ALIA, and THEIA. This, in tandem with indirect detection experiments, provides a multi-messenger test of our scenario. We see in Fig.~\ref{fig:multi-messenger_parameter_space} that almost the entire region of the $\langle \sigma v \rangle-m$ plane that is allowed by the FERMI-LAT constraints can be probed by GW detectors. Future indirect searches like CTA can access the bulk of this region that include masses between the weak scale and $10^4$~GeV.         

The crucial ingredient of our scenario is a strongly FOPT at temperatures below a few TeV which gives rise to DM mass. We largely focus on general conditions for its success and prospects of its multi-messenger probes. A natural direction for future investigations involves model-specific studies. Detection of GWs combined with a signal from indirect DM searches can help us reconstruct the potential governing the dynamics of the FOPT. This can then be used to build explicit particle physics models employing a FOPT to give mass to the DM. 


\begin{acknowledgments}
The work of R.A. is supported in part by NSF Grant No. PHY-2210367. The work of C.H. and P.H. is supported by the National Science Foundation under grant number PHY-2112680. 
\end{acknowledgments}

\begin{appendices}
\section*{Appendix}

\subsection{Freeze-Out of Cold Relics}
\label{appendix:freeze-out}

The Boltzmann equation reveals three processes responsible for the changing DM number density. Cosmic expansion and self-annihilation deplete the number density at rates of $3Hn$ and $\langle \sigma v \rangle n^2$, respectively. Equilibrium interactions at high temperatures help maintain the number density at a rate $\langle \sigma v \rangle n_\mathrm{eq}^2$. To factor out effects of cosmic expansion, one can define the yield of DM as $Y \equiv n / s$ where $s$ is the entropy density of the universe. During cosmic expansion $n$ and $s$ scale equivalently with $a^{-3}$, allowing Eq.~\ref{eq:Boltzmann_equation} to be rewritten as
\begin{equation}
    \label{eq:Boltzmann_equation_abundance_time}
    \dv{Y}{t} = - \langle \sigma v \rangle s \left[ Y^2 - Y_\mathrm{eq}^2 \right] \, ,
    \quad \quad
    Y \equiv n/s
\end{equation}
with $Y_\mathrm{eq} \equiv n_\mathrm{eq} / s$. One can further adimensionalize this equation with the parameterization of $x \equiv m / T$. Assuming the DM to be in thermal equilibrium with a radiation-dominated universe, the Friedmann equations give $\dd  x / \dd  t = Hx$ and Eq.~\ref{eq:Boltzmann_equation_abundance_time} becomes
\begin{equation}
    \label{eq:Boltzmann_equation_abundance}
    \dv{Y}{x} = - \frac{\langle \sigma v \rangle s}{Hx} \left[ Y^2 - Y_\mathrm{eq}^2 \right] \, ,
    \quad \quad
    x \equiv \frac{m}{T} \, .
\end{equation}
The Friedmann equations also yield the following expressions for the Hubble parameter and entropy density:
\begin{equation}
    \label{eq:radiation_domination}
    H^2 = \frac{8 \pi}{3 M_\mathrm{Pl}^2} g_*(T) \frac{\pi^2}{30} T^4 \, ,
    \quad \quad
    s = \frac{2\pi^2}{45} g_{*s}(T) T^3
\end{equation}
with Planck mass $M_\mathrm{Pl} = 1.22 \times 10^{19}$~GeV. In these expressions $g_*$ and $g_{*s}$ give the total degrees of freedom in the universal plasma and entropy density, respectively~\cite{Borsanyi:2016ksw}. This study approximates these degrees of freedom to be equal: $g_* = g_{*s}$. The sum of Standard Model and DM relativistic degrees of freedom gives the total number of degrees of freedom in the plasma, that is $g_* = g_\mathrm{SM} + g$.

\begin{figure}
    \centering
    \includegraphics[scale=0.172]{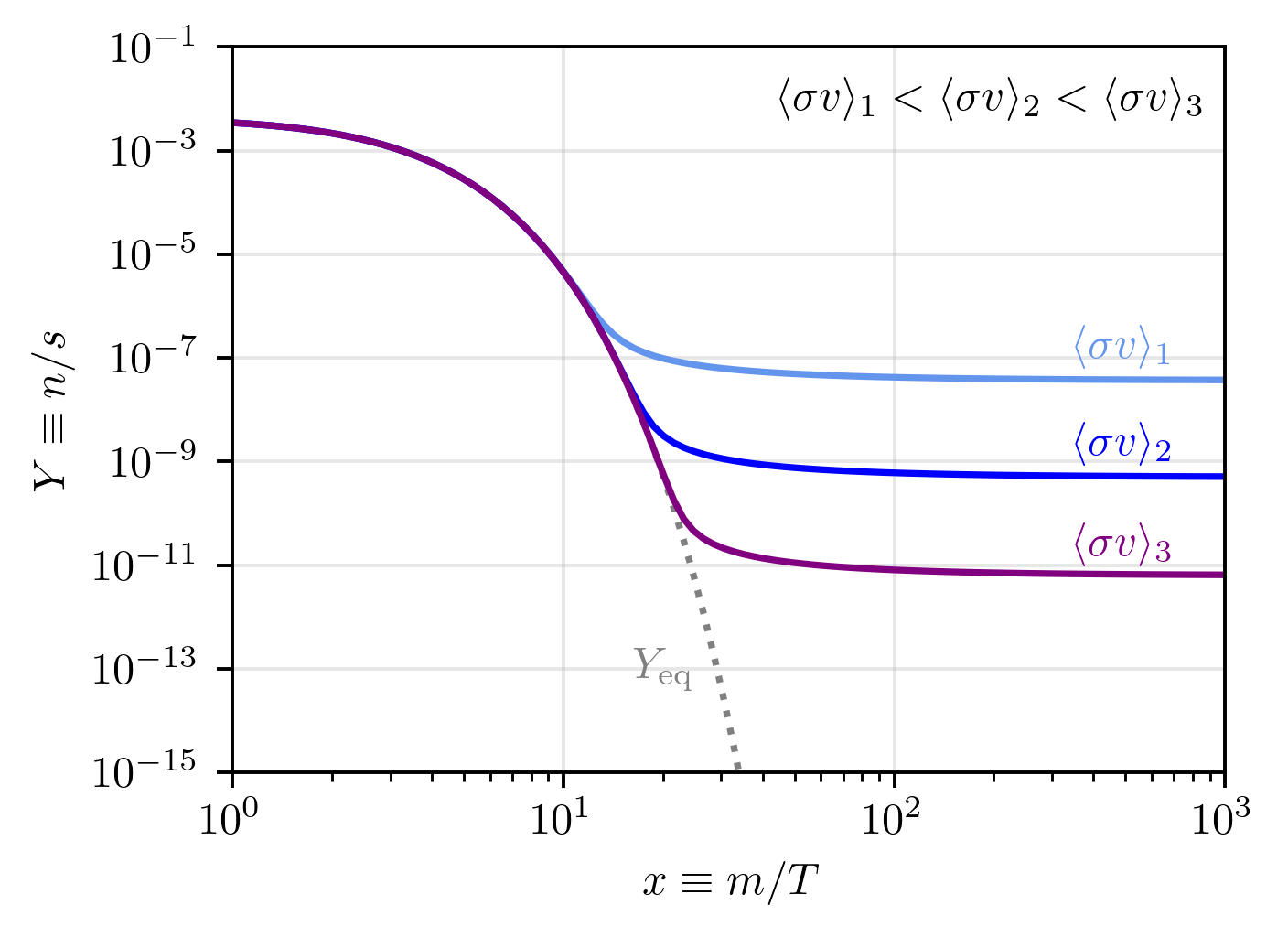}
    \caption{Typical solutions to Eq.~\ref{eq:Boltzmann_equation_abundance} with the boundary condition $Y(1) = Y_\mathrm{eq}(1)$. Departure from $Y_{\rm eq}$ marks the onset of freeze-out, which occurs at later times for higher $\langle \sigma v \rangle$.}
    \label{fig:comoving_abundance}
\end{figure}

Typical solutions of the Boltzmann equation are displayed in Fig.~\ref{fig:comoving_abundance} where $Y = Y_\mathrm{eq}$ is taken as a boundary condition for early times (small $x$). The universe is dense and hot throughout these times, keeping the DM yield near its thermal equilibrium value $Y_\mathrm{eq}$, while self-annihilation decreases the number of DM particles. Eventually the expansion of the universe ``freezes out" DM annihilation processes in that the number of DM particles within a comoving volume becomes too low for annihilation rates to significantly affect $Y(x)$. The approximate time at which this occurs is denoted by $x_f$. With a boundary condition specified for some unique Boltzmann equation, a characteristic freeze-out for the system is determined. For $x > x_f$, the DM comoving abundance is said to be ``frozen in" as the number of DM particles per comoving volume remains relatively constant after freeze-out. As long as freeze-out occurs well into the past, the frozen-in DM abundance determines its relic abundance as measured today: $\Omega_\mathrm{DM} h^2 = 0.12$~\cite{Planck:2018vyg}. Here, $h$ is determined by measurement of today's value of the Hubble parameter $H_0 = 100 \, h \,\, \mathrm{km} \, \mathrm{s}^{-1} \, \mathrm{Mpc}^{-1}$. With knowledge of the frozen-in DM abundance, we expect to find
\begin{equation}
    \label{eq:DM_relic_density}
        \Omega_\mathrm{DM} = \frac{m Y(x_\mathrm{today}) s_0}{\rho_{c, 0}} \, 
\end{equation}
for nonrelativistic DM. Although Eq.~\ref{eq:Boltzmann_equation_abundance} has no known closed-form solutions, one can find approximations for today's yield $Y(x_\mathrm{today})$ for two types of DM relics. Hot relics are models of DM wherein freeze-out occurs while the species is still relativistic, that is $x_f \ll 3$. Models with nonrelativistic DM at freeze-out ($x_f \gg 3$) are known as cold relics. The next section follows a standard analysis that can be found in Ref.~\cite{Kolb:1990vq}, for example.

The first simplifying assumption is made on the thermally averaged annihilation cross section $\langle \sigma v \rangle$. It is expected that $\sigma v \propto v^{k/2}$ where $k=0$ corresponds to $s$-wave annihilation and $k=1$ corresponds to $p$-wave annihilation. As $\langle v \rangle \propto T^{1/2}$, we parameterize the cross section as
\begin{equation}
\begin{split}
    \label{eq:cross_section_parameterization}
    \langle \sigma v \rangle &\equiv \sigma_0 (T/m)^k \\
    &= \sigma_0 x^{-k} \,\,
    \begin{cases}
        k = 0 & \Longleftrightarrow s\mathrm{-wave \,\, annihilation} \\
        k = 1 & \Longleftrightarrow p\mathrm{-wave \,\, annihilation}
    \end{cases} \, .
\end{split}
\end{equation}
Our study assumes $s$-wave annihilation is the dominant contribution and thus $k=0$. This parameterization allows Eq.~\ref{eq:Boltzmann_equation_abundance} to be rewritten as
\begin{equation}
    \label{eq:Boltzmann_equation_cold_relic}
    \begin{split}
    \dv{Y}{x} &= - \lambda x^{-k-2} \left[ Y^2 - Y_\mathrm{eq}^2 \right] \, , \\
    \lambda \equiv \frac{x \sigma_0 s}{H}
    &= \left[\frac{\sigma_0 s}{H} \right]_{x=1} 
    = \sqrt{\frac{\pi}{45}} \frac{g_{*s}}{\sqrt{g_*}} M_\mathrm{Pl} m \sigma_0
    \end{split}
\end{equation}
where a new parameter $\lambda$ has been defined which is approximately independent of $x$.
To track the yield's deviation from its equilibrium value, we express the right-hand side of Eq.~\ref{eq:Boltzmann_equation_cold_relic} in terms of $\Delta \equiv Y - Y_\mathrm{eq}$ to find
\begin{equation}
    \label{eq:Boltzmann_equation_Delta}
    \dv{Y}{x} = -\lambda x^{-k-2} \Delta (2 Y_\mathrm{eq} + \Delta) \, .
\end{equation}
As $Y_\mathrm{eq}$ decreases exponentially with $x$, we expect $\Delta \approx Y$ with cold relics for times after freeze-out, meaning Eq.~\ref{eq:Boltzmann_equation_Delta} is approximately
\begin{equation}
    \label{eq:Delta_equation_later_times}
    \dv{\Delta}{x} \approx -\lambda x^{-k-2} \Delta^2
    \quad \quad
    (x \gg x_f) \, .
\end{equation}
Despite the validity of this approximation holding for $x \gg x_f$, the standard method is to integrate the above equation from $x_f$ out to some $x \gg x_f$ to find
\begin{equation}
    \label{eq:integration}
    \begin{split}
        \int_{\Delta(x_f)}^{\Delta(x)}\frac{\dd \Delta}{\Delta^2} &\approx -\lambda \int_{x_f}^{x} \frac{\dd x^\prime}{{x^\prime}^{k+2}} \\
        \left[ \frac{1}{\Delta(x)} - \frac{1}{\Delta(x_f)} \right] &\approx \frac{-\lambda}{k+1} \left[ \frac{1}{x^{k+1}} - \frac{1}{x_f^{k+1}} \right] \\
        \Delta(x) &\approx \frac{k+1}{\lambda} \left[ \frac{1}{x_f^{k+1}} - \frac{1}{x^{k+1}} \right]^{-1} \\
        \Delta(x) &\approx \frac{k+1}{\lambda} x_f^{k+1} \, .
    \end{split}
\end{equation}
The third line of Eqs.~\ref{eq:integration} ignores $1/\Delta(x_f)$ since the yield at freeze-out is typically larger than its value at some later time. The last line utilizes $x \gg x_f$ and rids the dependence on $x$. With $\Delta(x) \approx Y(x)$, the standard approximation for today's DM yield is finally given by
\begin{equation}
    Y_\mathrm{st}(x_\mathrm{today}) \approx \frac{k+1}{\lambda} x_f^{k+1} \, .
\end{equation}
If instead the DM is produced after its characteristic freeze-out time, we can simply integrate Eq.~\ref{eq:Delta_equation_later_times} from a lower bound given by the time of production $x_P$. With $x_{\rm today} \gg x_P > x_f$, a similar result is found:
\begin{equation}
    Y(x_\mathrm{today}) \approx \frac{k+1}{\lambda} x_P^{k+1} \, .
\end{equation}

\subsection{Thermally Averaged M{\o}ller Speed}
\label{appendix:Moller_speed}
The M{\o}ller speed between two particles with four-momenta $p_1^\mu = (E_1, \mathbf{p}_1)$ and $p_2^\mu = (E_2, \mathbf{p}_2)$ is
\begin{equation}
    v \equiv \frac{\sqrt{(p_1^\mu p_{2 \mu})^2 - (m_1 m_2)^2}}{E_1 E_2}
\end{equation}
where $m_i^2 \equiv p_i^\mu p_{i \mu}$. Consider a colinear collision between particles of the same mass: $m_1 = m_2 \equiv m$. Furthermore, in taking a thermal average, we should expect each particle to have the same energy and momentum (although antiparallel), which we will call $E$ and $|\mathbf{p}|$. We then find
\begin{equation}
\begin{split}
    \langle v \rangle^2 &= \frac{(E^2 + |\mathbf{p}|^2)^2 - m^4}{E^4} \\
    &= 1 + \left( \frac{|\mathbf{p}|}{E} \right)^4 + 2\left( \frac{|\mathbf{p}|}{E} \right)^2 - \left( \frac{m}{E} \right)^4 \\
    &= 1 + \left( \frac{3T}{\gamma m} \right)^2 + 2\left( \frac{3T}{\gamma m} \right) - \frac{1}{\gamma^4} \\
    &= \frac{6}{x} \sqrt{1-v_{\rm rel}} + \frac{9}{x^2} (1-v_{\rm rel}) + v_{\rm rel}^2 (2 - v_{\rm rel}^2) \, .
\end{split}
\end{equation}
The relativistic equipartition theorem of $3T = |\mathbf{p}|^2 / E$ (along with $E = \gamma m$) was used in the third line, and $\gamma \equiv 1 / \sqrt{1-v_{\rm rel}}$ is the Lorentz factor between the two rest frames of the particles. We thus see that for cold relics with $x_f \gg 3$ and nonrelativistic relative speeds $v_{\rm rel} \ll 1$, the M{\o}ller speed at freeze-out is $\langle v \rangle \approx \sqrt{6/x_f}$.

\subsection{Scalar Decay}
\label{appendix:scalar_decay}

Consider scalar DM particles $\Delta$ produced inside the bubble walls of the developing broken phase via decay of a symmetry-breaking scalar field $\phi$. For scalar DM, the relevant terms in the Lagrangian density are 
\begin{equation}
    \label{eq:scalar_decay_Lagrangian}
    \mathcal{L} \supset - \frac{1}{2} m_\Delta^2 \Delta^2 - \frac{1}{2} \lambda \Delta^2 |\phi|^2
\end{equation}
The interaction term develops a three-point vertex when $\phi$ acquires a nonzero VEV, inducing a phase transition which we take to be of first order. Letting $\phi \rightarrow \phi + \langle \phi \rangle$, the interaction term becomes
\begin{equation}
    \label{eq:scalar_decay_term}
    \frac{1}{2} \lambda \Delta^2 |\phi + \langle \phi \rangle|^2 = 
    \frac{1}{2} \lambda \Delta^2 \left( |\phi|^2 + \langle \phi \rangle^2 + 2 \langle \phi \rangle |\phi| \right) \, .
\end{equation}
Here, the first term describes a quartic interaction between the DM and scalar as in the original Lagrangian, the second term provides the DM with additional mass, and the third term is a trilinear coupling between DM and scalar. This last term is responsible for the decay of $\phi$ into DM within bubble walls where $\langle \phi \rangle \neq 0$.

In this mechanism, the abundance of DM in the true vacuum phase is determined by the $\phi \rightarrow \Delta$ decay process, as well as how many scalars are able to penetrate the bubble walls. Suppose the scalar field has some yield $Y^\phi_\mathrm{fv} = n^\phi_\mathrm{fv} / s$ and decays into $N$ DM particles with probability $|\mathcal{T}_{\phi \rightarrow N \Delta}|^2$ at any given time. If some fraction $Y^\phi / Y^\phi_\mathrm{fv}$ of scalar particles penetrate the bubble walls into the true vacuum, the yield of DM in the true vacuum is given by
\begin{equation}
    Y = \frac{Y^\phi}{Y^\phi_\mathrm{fv}} \frac{n^\phi(T_\mathrm{PT})}{s(T_\mathrm{PT})} N |\mathcal{T}_{\phi \rightarrow N \Delta}|^2 \, .
\end{equation}
Prior to the FOPT, we assume the scalar field $\phi$ to be in thermal equilibrium with the field content of the universe so that $n^\phi(T) = \zeta(3) \gamma_w g_\phi T^3 / \pi^2$ which gives
\begin{equation}
    \label{eq:DM_abundance_scalar_decay}
    Y = \frac{Y^\phi}{Y^\phi_\mathrm{fv}} \frac{45 \zeta(3) g_\phi}{2\pi^4 g_{*,s}} N |\mathcal{T}_{\phi \rightarrow N \Delta}|^2 \, .
\end{equation}
To meet the condition of DM abundance to be enhanced from its standard freeze-out value, we require $Y \geq Y_c(T_\mathrm{PT})$. Combining this condition with Eqs.~\ref{eq:critical_density} and \ref{eq:DM_abundance_scalar_decay} yields
\begin{equation}
    \label{eq:abundance_lower_bound_scalar_decay}
    \frac{Y^\phi}{Y^\phi_\mathrm{fv}} \langle \sigma v \rangle T_\mathrm{PT} \geq
    \frac{2\sqrt{6} \pi^{5/2}}{\zeta(3) M_\mathrm{Pl}} \frac{\sqrt{g_*}}{g_\phi} \frac{1}{N |\mathcal{T}_{\phi \rightarrow N \Delta}|^2} \, .
\end{equation}

\end{appendices}

\bibliography{main}
\end{document}